\documentstyle[aps,pra,twocolumn]{revtex}
\begin{document}
\draft
\twocolumn[\hsize\textwidth\columnwidth\hsize\csname @twocolumnfalse\endcsname
\title{DIMENSIONAL CROSSOVER IN HEAVY FERMIONS}
\author{Mucio A.Continentino}
\address{Instituto de Fisica,
Universidade Federal Fluminense \protect\\
Campus da Praia Vermelha, Niter\'oi, 24.210-340, RJ, Brasil}
\date{\today}
\maketitle
\begin{abstract}
Recently we have shown that a one-parameter scaling, $T_{coh}$, describes the physical behavior of several  heavy fermions in a region of their phase diagram. In this paper we fully characterize this region, obtaining  the uniform susceptibility, the resistivity and the specific heat. This allows for an explicit evaluation of the Wilson and the Kadowaki-Woods ratios in this regime. These quantities turn out to be independent of the distance $|\delta|$ to the critical point. The theory of the one-parameter scaling  corresponds to a zero dimensional approach. Although spatial correlations are irrelevant in this case, time fluctuations are critically correlated and the generalized hyperscaling relation is satisfied for $d=0$.  The crossover from $d=0$ to $d=3$ is smooth. It occurs at a lenght scale which is inversely related to the {\em stiffness} of the lifetime of the spin fluctuations. 

\end{abstract}

] \newpage

\section{Introduction}

Most of the physical properties of heavy fermions can be attributed to the
fact that these systems are close to a quantum critical point (QCP) \cite{mucio0}. The
critical point arises as a competition between Kondo effect and magnetic
order induced by RKKY coupling. In the non-critical side of the phase
diagram, where the system never orders, a scaling approach reveals the
existence of a new characteristic temperature, the {\em coherence temperature%
} $T_{coh}\propto |\delta |^{\nu z}$, below which the system exhibits Fermi
liquid behavior \cite{mucio1}. In this equation, $|\delta |=|J_Q-J_Q^c|$ measures the
distance to the $T=0$ critical point and $\nu $ and $z$ are respectively the
correlation length and dynamic critical exponents. $J_Q$ is the coupling
between the local moments and $J_Q^c$ its critical value, at which the
magnetic instability characterized by the wavevector $Q$ occurs. At the QCP,
i.e., $|\delta |=0$, the system does not cross the {\em coherence line} and
consequently exhibits {\em non-Fermi liquid behavior} down to $T=0$ \cite{mucio0}. 

Recently we have shown that a one-parameter
scaling, the coherence temperature, is able to describe the pressure
behavior of several physical quantities for different heavy fermions \cite{mucio0}. In this report we show this is due to the
flatness of the spectrum of spin fluctuations. We use the spin fluctuation theory of a nearly
antiferromagnetic system \cite{moriya} \cite{taki} to fully characterize this one-parameter scaling regime and calculate the
specific heat, the uniform susceptibility and the resistivity. This allows for an explicit evaluation of the {\em Wilson ratio} and
the {\em Kadowaki-Woods ratio} \cite{kado} between the coefficient of the $T^2$ term in
the resistivity and that of the linear term of the specific heat.
 
We make use of the spin fluctuation model since it is  a Gaussian theory and for the problem considered here, where the effective dimension $d_{eff} = d+z > d_c =4$ with $d_c$ the upper critical dimension for the magnetic transition, it gives the {\em correct} description of the quantum critical behavior \cite{mucio0} \cite{millis}.

\section{Specific heat}

We start form the expression for the specific heat given by the
spin-fluctuation theory for a nearly antiferromagnetic electronic system \cite{taki}. We
will use here the notation of Ref. \cite{taki}, 
\begin{equation}
C/T=\frac{\partial ^2}{\partial T^2}\left[ \frac 3\pi \sum_{\vec{q}%
}T\int_0^\infty d\lambda \frac{d\lambda }{e^\lambda -1}\tan ^{-1}\left( 
\frac{\lambda T}{\Gamma _q}\right) \right] 
\end{equation}
where 
\[
\Gamma _q=\Gamma _L(1-J_Q\chi _L)+\Gamma _L\chi _LAq^2
\]
$\Gamma _L$ and $\chi _L$ are local parameters defined through the local
dynamical susceptibility \cite{taki} 
\[
\chi _L(\omega )=\frac{\chi _L}{1-\omega /\Gamma _L}
\]
$J_{Q\text{ }}$ as before is the q-dependent exchange coupling between f-moments 
and $A$ is the {\em stiffness} of the lifetime of the spin fluctuations defined by the small
wavevector expansion of the magnetic coupling close to the wavevector $Q$,
i.e., $J_Q-J_{Q+q}=Aq^2+\cdots $. Then $\Gamma _q$ can be rewritten as 
\[
\Gamma _q=\Gamma _L\chi _LA\xi ^{-2}[1+q^2\xi ^2]
\]
where the correlation length $\xi =\left( A/|J_Q-J_Q^c|\right) ^{1/2}$
diverges at the critical value of the coupling $J_Q^c=\chi _L^{-1}$ with the Gaussian 
critical exponent $\nu =1/2$. Consequently we have for the specific heat 
\[
C/T=\frac{\partial ^2}{\partial T^2}\{\frac 3\pi \sum_{\vec{q}%
}T\int_0^\infty d\lambda \frac{d\lambda }{e^\lambda -1}
\]
\begin{equation}
\tan ^{-1}\left( \frac{\lambda T\xi ^z}{\Gamma _L\chi _LA(1+q^2\xi ^2)}%
\right) \}
\end{equation}
where the dynamic critical exponent $z=2$, typical of antiferromagnetic spin fluctuations. The exponential cuts off the
contribution for the integral from large values of $\lambda $, consequently
for $(T\xi ^z/\Gamma _L\chi _LA)<<1$ we can expand the $\tan ^{-1}$ for
small values of its arguments. The above condition can be written as $%
T<<T_{coh}$, where the coherence temperature 
$$k_B T_{coh}=\Gamma _L\chi _L|J_Q-J_Q^c|\propto |\delta|^{\nu z}$$
is {\em independent} of $A$ and $\nu z=1$. In this regime the system shows Fermi liquid behavior and we get 
\begin{equation}
C/T=\frac{\partial ^2}{\partial T^2}\left[ \frac{3T^2\xi ^z}{\pi \Gamma
_L\chi _LA}\int_0^\infty \frac{d\lambda \lambda }{e^\lambda -1}\sum_{\vec{q}}%
\frac 1{1+q^2\xi ^2}\right] 
\end{equation}
Changing the $\sum_{\vec{q}}$ into an integral we find (d=3), 
\begin{equation}
C/T=\frac{\partial ^2}{\partial T^2}\left[ \frac{\pi T^2\xi ^{(z-d)}}{%
2\Gamma _L\chi _LA}\frac{4\pi V}{(2\pi )^3}\int_0^{q_c\xi }\frac{dyy^2}{1+y^2%
}\right] 
\end{equation}
which yields 
\begin{equation}
C/T=\frac{\pi \xi ^{(z-d)}}{\Gamma _L\chi _LA}\frac{4\pi V}{(2\pi )^3}q_c\xi
\left( 1-\frac{\tan ^{-1}q_c\xi }{q_c\xi }\right) 
\end{equation}
Taking the limit $q_c\xi <<1$ and since $tan^{-1}y\approx
y-y^3/3+y^5/5+\cdots $ for small $y$, we get \cite{mucio0} 
\begin{equation}
C/T=\frac{\pi \xi ^{(z-d)}}{\Gamma _L\chi _LA})\frac{4\pi V}{(2\pi )^3}%
q_c\xi \left[ \frac 13(q_c\xi )^2-\frac 15(q_c\xi )^4+\cdots \right] 
\end{equation}
The first term is independent of $A$ and yields essentially the result of
the {\em local interacting model} \cite{lacroix}, i.e., 
\begin{equation}
C/T=\frac{\pi Nk_B^2}{\Gamma _L\chi _L}\frac 1{|J_Q-J_Q^c|}
\end{equation}
In fact this  could have been obtained directly from Eq.3, neglecting
the $q$-dependence of $\Gamma _q$ and with $\sum_{\vec{q}}\rightarrow N$ \cite{lacroix}. In
the equation above the correct units have been restored. Note that the limit 
$q_c\xi <<1$ may be written as $q_c\left( A/|J_Q-J_Q^c|\right) ^{1/2}<<1$.
This can be satisfied either because the system is far away from the
critical point, i.e., $|J_Q-J_Q^c|$ is large, or because $A$ is small. If
we write the condition $q_c\xi <<1$ in the form $q_c/\sqrt{|J_Q-J_Q^c|}<<1/%
\sqrt{A}$ , we notice that when $A\rightarrow 0$ this condition becomes valid {\em %
arbitrarily close} to the quantum critical point. 

We can  rewrite the equation above for the specific heat as $C/T = \pi Nk_B/T_{coh}$. The large effective masses of heavy fermions are then related to the smallness of $T_{coh}$ consistent with the experimental observations.

\section{Susceptibility and Wilson ratio}

The zero temperature uniform susceptibility of the nearly antiferromagnetic
system in the limit $q_c\xi <<1$ can be directly obtained from the magnetic
field $(h)$ dependent, $T=0$, q-independent free energy \cite{mucio0} \cite{sachdev}, 
\begin{equation}
f=-\frac{3N}{2\pi }\int_0^{\omega _c}d\omega \tan ^{-1}\left[ \frac{\omega +h%
}{\Gamma _L\chi _L|J_Q-J_Q^c|}\right]
\end{equation}
Integrating, differentiating once, twice, taking the value at $h=0$ and the limit $\omega
_c\rightarrow \infty $ we obtain 
\begin{equation}
\chi _0= - \left( \frac{\partial ^2f}{\partial h^2}\right) _{h=0}=\frac{3N\mu ^2%
}{2\pi \Gamma _L\chi _L}\frac 1{|J_Q-J_Q^c|}
\end{equation}
or $\chi _0= 3N\mu ^2/2\pi T_{coh}$. 

The Wilson ratio is given by 
\[
\frac{\chi _0/ \mu ^2}{C/\pi^2 k_B^2T}= \frac {3}{2} = 1.5
\]
which turns out to be a universal number since the dependence on the
distance to the critical point, $|J_Q-J_Q^c|$ and on the dimensionless quantity $%
\Gamma _L\chi _L$ cancels out. This ratio can increase if we decrease the
energy cut-off of the excitations contributing to the specific heat. We
emphasize that the above result is valid in the regime $q_c\xi <<1$, that
is, if the system satisfies the condition, $q_c/\sqrt{|J_Q-J_Q^c|}<<1/\sqrt{%
A}$.

\section{Resistivity and Kadowaki-Woods ratio}

The resistivity due to spin fluctuations in the regime $q_c\xi <<1$ is given
by \cite{mucio0} \cite{lederer} 
\begin{equation}
\rho =\rho _0\frac 1T\int_0^\infty d\omega \frac{\omega \Im m\chi _Q(\omega )%
}{(e^{\beta \omega }-1)(1-e^{-\beta \omega })}
\end{equation}
where 
\[
\Im m\chi _Q(\omega )=\chi _Q^s\frac{\omega \xi _L^z}{1+(\omega \xi _L^z)^2} 
\]
with 
\[
\chi _Q^s=\frac 1{|J_Q-J_Q^c|} 
\]
and 
\[
\xi _L^z=\frac{\chi _Q^s}{\Gamma _L\chi _L} 
\]
The quantity $\rho _0$ is given by, 
\[
\rho _0=(\frac JW)^2\frac{m_c}{n_ce^2\tau _{Fc}} (n/n_c)
\]
where $J$ is the coupling constant per unit cell between localized and
conduction electrons. $W$, $m_c$ and $n_c$ are the bandwidth, the mass and
the number of conduction electrons per unit volume with Fermi momentum $%
k_{Fc}$, such that $\hbar {\tau _{Fc}}^{-1}=\hbar ^2k_{Fc}^2/2m_c$. $n$ is the number of atoms per unit volume.

Using the definitions above, we can rewrite the resistivity as, $\rho =\rho
_0\Gamma _L\chi _LR(\tilde{T})$ , where 
\begin{equation}
R(\tilde{T})=\frac 1{\tilde{T}}\int_0^\infty d\tilde{\omega}\frac 1{(e^{%
\tilde{\omega}/\tilde{T}}-1)(1-e^{-\tilde{\omega}/\tilde{T}})}\frac{\tilde{%
\omega}^2}{1+\tilde{\omega}^2}
\end{equation}
with $\tilde{\omega}=\omega \xi _L^z$ and $\tilde{T}=T\xi _L^z$. For $%
T<<T_{coh}$ we have $R(T<<T_{coh})\approx \frac{\pi ^2}3(\frac T{T_{coh}})^2$
and finally 
\[
\rho (T<<T_{coh})=\rho _0\Gamma _L\chi _L\frac{\pi ^2}3(\frac T{T_{coh}}%
)^2=A_RT^2 
\]
where 
\[
A_R=\frac{\rho _0\pi ^2}3\frac{k_B^2}{\Gamma _L\chi _L}\frac 1{|J_Q-J_Q^c|^2} 
\]
The Kadowaki-Woods ratio \cite{kado} $A_R/({C/T})^2$ is given by 
\[
\frac{A_R}{(C/T)^2}=\frac{\rho _0\Gamma _L\chi _L}{3(Nk_B)^2} 
\]
which depends on the local parameters, $\Gamma _L\chi _L$, consequently on
the nature of the magnetic ion ($f$ or $d$, for example), but{\em \ not} on
the distance to the critical point, $|J_Q-J_Q^c|$.  
We get $A_R/(C/T)^2 \approx  4.8x10^{-9} \rho_0 \Gamma_L \chi_L (moleK/mJ)^2$. Using the $T=0$ value of $\Gamma _L\chi _L = 1/2 \pi$ \cite{taki} and the experimental value for this ratio  we can find  $\rho_0$ and determine microscopic parameters of the system \cite{coqblin}.
 
We point out that in the
q-dependent regime, $q_c\xi \geq 1$, also $\rho = A_R^M T^2$    at low temperatures but the coefficient 
$A_R^M \propto |J_Q-J_Q^c|^{-1/2}$ \cite{taki} and consequently does not scale as $T_{coh}^{-2}$ , in  disagreement with experiments in heavy fermions \cite{mucio0}.

\section{The non-Fermi liquid regime}

As the system gets close to the QCP and $q_c \xi \ge 1$,  the system should be described using the full q-dependent
susceptibility. In particular at the quantum critical point, i.e.,  $|\delta |=0$ but finite
temperatures, the neglect of the q-dependence of $\chi(q, \omega)$ leads to unphysical behavior as, a
constant resistivity and diverging specific heat. The spin fluctuation
theory predicts that the Neel line close to the quantum critical point
behaves as \cite{millis} $T_N\propto |\delta |^\psi $, where the shift exponent $\psi
=z/(d+z-2)=2/3 \neq \nu z=1$. The appropriate generalized scaling form of the free energy for this case is given by \cite{mucio0}, $f\propto {|\delta (T)|}^{2-\alpha }F_c[t]$ with
$t=T/{|\delta (T)|}^{\nu z}$
and $\delta (T)=\delta (T=0)-uT^{1/\psi }$
where $u$ is a constant \cite{millis1}. The singularities along the Neel line, ${|\delta
(T)|}=0$, are described by  {\em tilde} exponents $\tilde{\alpha}
$, $\tilde{\nu}$, etc., different from those associated with the zero
temperature fixed point (the {\em non-tilde} exponents) \cite{mucio1}. The scaling
function $F_c[t=0]=$ $constant$ and $F_c[t\rightarrow \infty ]\propto t^x$
with $x=(\tilde{\alpha}-\alpha )/\nu z$ such that close to the critical Neel
line we obtain the correct asymptotic behavior, $f\propto A(T){|\delta (T)|}%
^{2-\tilde{\alpha}}$, where the amplitude $A(T)=T^{\frac{\tilde{\alpha}%
-\alpha }{\nu z}}$ \cite{mucio1}. For the
specific heat we find
\begin{equation}
C/T\propto u^{2- \tilde{\alpha}}T^{\frac{(2-\tilde{\alpha})(\nu z-\psi )+\nu \psi (d-z)}{\nu
z\psi }}
\end{equation}
Assuming thermal Gaussian exponents, essentially $\tilde{\alpha}=1/2$, we
get, $C/T\propto u^{3/2}T^{5/4}$ for $\psi =2/3$, $\nu =1/2$ and $z=2$, instead of $%
C/T\propto T^{1/2}$ for the case of extended scaling ($\nu z = \psi$) \cite{taki}. The staggered
susceptibility $\chi _Q(\delta =0,T)\propto T^{-\tilde{\gamma}/\psi
}=T^{-3/2}$ since $\gamma =\tilde{\gamma}=1$ \cite{moriya}. For the
correlation length we get $\xi^{-2} \propto uT^{ \frac{d+z-2}{z}}$ since $\nu =\tilde{\nu}$.
This should be compared to Eq. 3.11 of Ref. \cite{millis}. Notice that the thermal and $T=0$
Gaussian  critical exponents are the same, except
for the exponent $\alpha $ of the free energy ($\tilde{\alpha}=2-\tilde{\nu} d=1/2$
, $\alpha =2- \nu(d+z)=-1/2$) and this is the reason our prediction for
the specific heat at $|\delta |=0$ is different from the spin fluctuation
result \cite{moriya}. Notice that while the Gaussian theory yields the correct exponents for the zero temperature transition, this is not the case for the finite temperature Neel transitions and the expression above for $C/T$ may be used with
non-Gaussian thermal exponents, for example, with those of the 3d Heisenberg
model.

\section{Conclusion}

We have calculated the Wilson and Kodawaki-Woods ratio for a model of nearly
antiferromagnetic systems in the regime $q_c\xi <1$. These
quantities turn out to be constants, i.e., independent of the distance $%
|\delta |$ to the critical point. The spin fluctuation theory in the regime $%
q_c\xi <1$ corresponds to a local interacting model and yields a one-parameter scaling  since $C/T\propto
T_{coh}^{-1},$ $\chi _0\propto T_{coh}^{-1},$ $A_R\propto T_{coh}^{-2}$ with 
$T_{coh}\propto \xi ^{-z}$. The local interacting model becomes valid arbitrarely close to the QCP as the {\em stiffness} $A$ of the lifetime of the spin fluctuations vanishes.  It can  be regarded as a zero dimensional
theory since, in spite that spatial correlations are irrelevant in this
regime, time fluctuations are critically correlated and the quantum
hyperscaling relation $2- \alpha = \nu (d+z)$ is satisfied for $d=0$ \cite{mucio0}. This is a direct consequence
of the quantum character of the transition. As the system gets closer to the critical point  and $q_c \xi >1$, the full q-dependence of the dynamic susceptibility must be taken into account. The crossover from $d=0$ to
three dimensional behavior is smooth. It occurs at a length scale which is
inversely related to the {\em stiffness} $A$. It is
possible that there is an intervening region dominated by two-dimensional
fluctuations \cite{rosch} before the system finally settles in three dimensional
criticality.


\begin{references}
\bibitem{millis1} In Ref. \cite{millis} $u$ is the dangerously irrelevant quartic interaction.
\end{references}
\end{document}